\documentclass{PoS}
\usepackage{graphicx}% Include figure files
\usepackage{enumerate}% For non-numerical items
\usepackage{bm}% bold math
\usepackage{slashed}
\usepackage{verbatim}
\usepackage{amsmath,amssymb}
\usepackage{epsfig}
\usepackage{hhline}
\usepackage{placeins}
\newcommand\mpi {m_{\pi}}

\newcommand\Eq[1]{Eq.~(\ref{eq:#1})}

\newcommand\Fig[1]{Fig.~\ref{fig:#1}}

\newcommand\beq{\begin{eqnarray}}
\newcommand\eeq{\end{eqnarray}}

\newcommand{\Tr}{{\rm Tr\,}}

\title{Baryon properties in meson mediums from lattice QCD}

\ShortTitle{Baryon properties in meson mediums from lattice QCD}

\author{\speaker{Amy Nicholson}%
         \\
        {Maryland Center for Fundamental Physics, Department of Physics,
University of Maryland, College Park MD 20742-4111, USA\\
        E-mail: \email{amynn@umd.edu}}}

\author{William Detmold\\
        Center for Theoretical Physics, Massachusetts Institute of Technology, Cambridge, MA 02139, USA\\
E-mail: \email{wdetmold@mit.edu}}

\abstract{We present results for the ground-state mass shifts of octet baryons due to the presence of a medium of pions or kaons from a lattice QCD calculation performed at a single value of the quark mass, corresponding to a pion mass of $m_{\pi} \sim$ 390 MeV, and a spatial volume $V \sim (4\mathrm{ fm})^3$. We use a canonical approach in which correlators are formed using a single baryon propagator and a fixed number of meson propagators, up to $n=9$. From the ground-state energies we calculate two- and three-body interaction parameters. We also extract combinations of low-energy constants by comparing our results to tree level chiral perturbation theory at non-zero isospin/kaon chemical potential.}

\FullConference{31st International Symposium on Lattice Field Theory LATTICE 2013\\
                 July 29 - August 3, 2013\\
                 Mainz, Germany}

\begin{document}

\section{Introduction}
While understanding the properties of systems of multiple hadrons from first principles is an important goal for the field of nuclear physics, our only reliable tool for performing QCD calculations in the low-energy regime, lattice QCD, suffers from severe issues with noise at finite baryon density \cite{Lepage:1989hd,Barbour:1986jf,Kogut:1994eq}. Multiple meson systems, on the other hand, do not pose the same problems, and have proven to be useful test beds for probing the limits of many-body calculations on the lattice \cite{Detmold:2008fn,Beane:2007es,Detmold:2008bw,Detmold:2012wc,Detmold:2012pi,Detmold:2011kw}. Furthermore, studying many-meson systems allows us to explore the phenomena of Bose Einstein condensation, and may prove essential for understanding the equation of state of neutron stars, as pion condensation \cite{Migdal,Sawyer:1972cq,Scalapino:1972fu,PhysRevLett.30.1340} and kaon condensation \cite{Kaplan:1986yq} have both been proposed to occur in their cores. 

To date, many-hadron lattice QCD calculations have been largely limited to pure mesonic systems, however, most phenomenologically relevant systems also contain baryons. In this work, we calculate the ground-state energies of four systems consisting of a single baryon and up to $n=9$ mesons: $\Xi^0\left(\pi^{+}\right)^n$, $\Sigma^{+}\left(\pi^{+}\right)^n$, $p\left(K^{+}\right)^n$, and $n\left(K^{+}\right)^n$. We include only a single baryon to render the signal-to-noise ratio manageable, and the systems are chosen to avoid possible annihilation between valence quarks. From the ground-state energies, we compute meson-baryon scattering lengths and meson-meson-baryon three-body interactions, and extract several combinations of low-energy constants (LECs) using results from tree-level heavy baryon chiral perturbation theory (HB$\chi$PT).

\section{Correlation functions}

To tackle the $N_u!N_d!N_s!$ contractions, where $N_i$ is the number of quarks of flavor $i$ in a given interpolating operator, we use a method based on the formalism presented in \cite{Detmold:2008fn,Beane:2007es,Detmold:2011kw}. In this formalism, the mesons are packaged into $12\times 12$ matrices, 
\beq
\Pi_{a, \alpha, b, \beta} &=& \sum_{c,\gamma} \sum_{\mathbf{x}}\left[ S_{q_1}(\mathbf{x},t;\mathbf{0},0) \gamma_5 \right]_{b,\beta,c,\gamma}\left[ S_{q_2}^{\dagger}(\mathbf{x},t;\mathbf{0},0) \gamma_5 \right]_{a,\alpha,c,\gamma} \to \Pi_{i,j}\ ,
\eeq
where $S_q(\mathbf{x},t;\mathbf{0},0)$ is the propagator for quark flavor $q$ from point $(\mathbf{0},0)$ to $(\mathbf{x},t)$, $(a,b,c)$ indices represent color, Greek indices represent spin, $(i,j)$ indices run over the 12 color/spin combinations, and the sum over spatial coordinates projects the meson onto zero momentum. 

For baryons having a single quark to be contracted with the medium, we form the following $12\times 12$ baryon matrices,
\beq
B_{a,\alpha,d,\rho}^{(\Xi,n)}= \sum_{b,\beta,c,\gamma,\lambda} B_{a,\alpha, b,\beta, c,\gamma,\lambda} \epsilon_{d,b,c}(C\gamma_5)_{\beta,\rho}(1\pm \gamma_4)_{\gamma,\lambda}\to B_{i,j}^{(\Xi,n)} \, ,\cr
B_{a,\alpha, b,\beta, c,\gamma,\lambda}= \sum_{\sigma,h,i,j}[S_{q_1}C\gamma_5]_{a,\alpha,h,\sigma}[S_{q_2}]_{b,\beta,i,\sigma}[S_{q_3}]_{c,\gamma,j,\lambda}\epsilon_{h,i,j} \, ,
\eeq
where $q_{1,2,3}$ are the quark flavors, $C$ is the charge conjugation matrix, all propagators are from $(\mathbf{0},0)$ to $(\mathbf{x},t)$, and $\mathbf{x}$ is summed over as in the mesonic case (spatial indices have been suppressed). In this form, the two quarks which are not contracted with the medium are partially contracted into a $\bar{3}$ color irrep with a single open spin index, effectively behaving as the antiquark in the meson blocks. Correlation functions are formed from these objects using the following relation,
\beq
\label{eq:DetExp}
\det(1 + \lambda \Pi + \kappa B^{(\Xi),n})=\frac{1}{12!}\sum_{j=1}^{12}\sum_{k=0}^{j}\left( \begin{array}{c}
12 \\
j \\
\end{array} \right)\left( \begin{array}{c}
j \\
k \\
\end{array} \right) \lambda^k \kappa^{j-k} C_{k,j-k}(t) = e^{\Tr[\log(1 + \lambda \Pi + \kappa K)]} \, .
\eeq 
Expanding the right-hand side of this equation, and collecting terms with $n$ powers of $\lambda$ and one power of $\kappa$ gives us the correlation function, $C_{n,1}(t)$ for the single baryon, $n$ meson system. 

For systems in which the baryon may interchange two quarks with the medium, we form the following $144\times 144$ matrix:
\beq
B_{d,\alpha,b,\rho,f,\lambda,c,\gamma}^{(\Sigma,p)}= \sum_{a,\beta,\sigma} B_{a,\alpha, b,\beta, c,\gamma,\lambda} \epsilon_{a,d,f}(C\gamma_5)_{\beta,\rho}(1\pm \gamma_4)_{\sigma,\lambda}\to B_{I,J}^{(\Sigma,p)} \, ,
\eeq
with indices $I,J$ which run over the spin/color of both quarks. We may now use \Eq{DetExp}, replacing $B^{(\Xi),n}\to B^{(\Sigma,p)}$, and $\Pi \to \Pi \otimes 1 + 1 \otimes \Pi$, representing outer products of the $\Pi$ matrices with the $12\times 12$ identity matrix. These methods greatly reduce computational time by converting an intractable number of index contractions into relatively few traces of matrix products, which may be computed once and reused to form all systems with up to 12 of any given quark propagator.

\section{Energy splittings}
To extract the ground state energies, correlation functions were computed on gauge field configurations produced by the Hadron Spectrum Collaboration (for details, see Ref.~\cite{Lin:2008pr}) using a $n_f=2+1$-flavor anisotropic dynamical tadpole-improved clover fermion action and a Symanzik-improved gauge action with $m_{\pi} \sim 386$ MeV and $m_K \sim 543$ MeV. The spatial lattice spacing for these ensembles is $b_s=0.1227(8)$, and the anisotropy parameter $\xi=b_{t}/b_s \sim 3.5$. We use ensembles with a large volume, ($32^3$) to ensure that we are near the scattering threshold, and a large temporal extent ($T=256$) to eliminate thermal effects. The quark propagators were computed by the NPLQCD collaboration (see Ref.~\cite{Beane:2009ky}), and were generated using the same fermion action as was used for gauge field generation. Details of the analysis of the correlation functions and numerical values for the energy splittings may be found in Ref.~\cite{Detmold:2013gua}.

\section{Scattering parameters}
Scattering parameters may be extracted using a generalization of L\"usher's formula \cite{Luscher:1986pf}, which relates the energy levels of two particles in a box to their scattering phase shifts. Equivalent formulations for multiple bosons have been computed in Ref.~\cite{Beane:2007qr,Detmold:2008gh} for identical bosons and in Ref.~\cite{Smigielski:2008pa} for two species of bosons to $\mathcal{O}(L^{-7})$ and $\mathcal{O}(L^{-6})$, respectively, in a perturbative expansion for large volumes. Because our systems contain a single fermion, spin statistics do not come into play and we may use the two species formulation, with the particles in the medium and the baryon acting as distinguishable species. The form we use for the energy splittings is,

\beq
\label{eq:LuscherExp}
    \Delta E_{MB}(n,L) = \frac{2{\pi}\bar{a}_{MB}n}{\mu_{MB}L^3}\left[1-\left(\frac{\bar{a}_{MB}}{\pi{L}}\right)\mathcal{I}+\left(\frac{\bar{a}_{MB}}{\pi{L}}\right)^{2}  \left(\mathcal{I}^2 +\mathcal{J}\left[-1+2\frac{\bar{a}_{MM}}{\bar{a}_{MB}}(n-1)\left(1+\frac{\mu_{MB}}{m_M}\right)\right.\right.\right. \cr
    \left.+\left(\frac{\bar{a}_{MB}}{\pi{L}}\right)^{3}\left(-\mathcal{I}^{3}
    +\sum_{i=0}^{2}\left(f^{\mathcal{I}\mathcal{J}}_{i}\mathcal{I}\mathcal{J}
    +f^{\mathcal{K}}_{i}\mathcal{K}\right)\left(\frac{\bar{a}_{MM}}{\bar{a}_{MB}}\right)^{i}\right)
    \right]+\frac{n(n-1)\bar{\eta}_{3,MMB}(L)}{2L^6}+\mathcal{O}(L^{-7}) \ ,\cr 
\eeq
where the effective scattering lengths, $\bar{a}_{MB}$ and $\bar{a}_{MM}$ are the inverse of the scattering phase shifts, $\left(p \cot \delta(p)\right)^{-1}$, corresponding to meson-baryon and meson-meson interactions, respectively, and the definitions of the mass-dependent coefficients $f_i$ are given in \cite{Detmold:2013gua}. The volume-dependent, but renormalization group invariant, three-body interaction, $\bar{\eta}_{3,MMB}(L)$, is shown explicitly in \cite{Detmold:2008gh}, as are the values for the geometric constants $\mathcal{I}$, $\mathcal{J}$, and $\mathcal{K}$.

\begin{figure}
\centering
\includegraphics[width=0.4\linewidth]{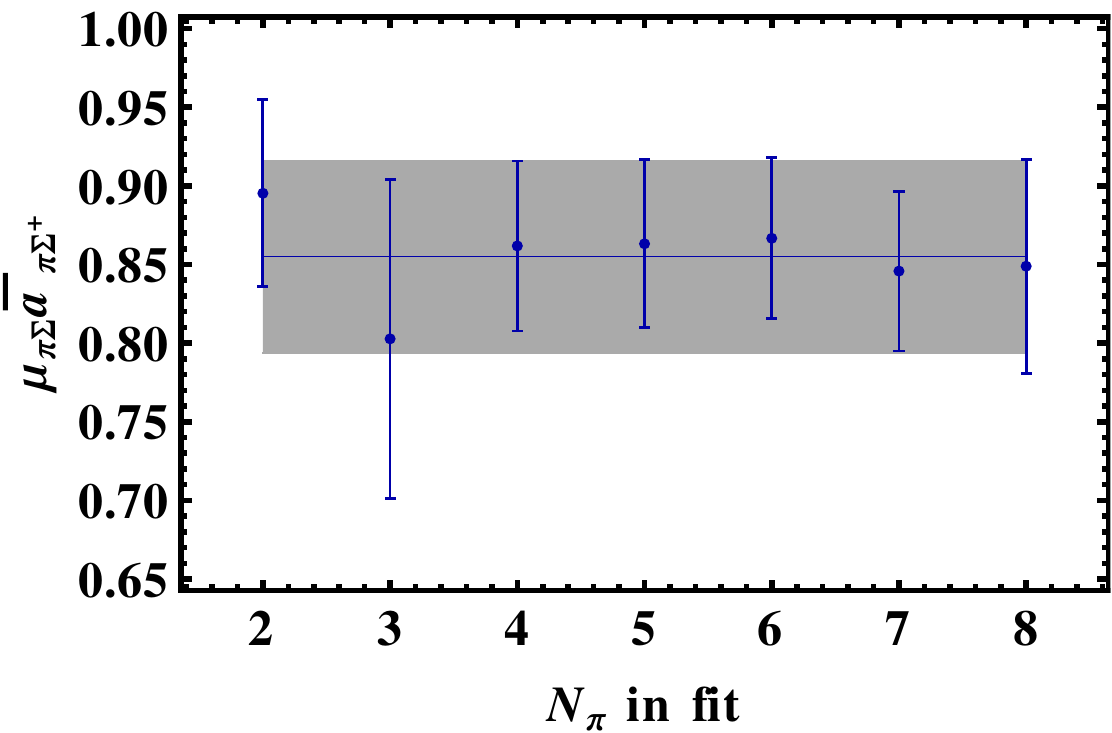}\hspace{2mm}
\includegraphics[width=0.4\linewidth]{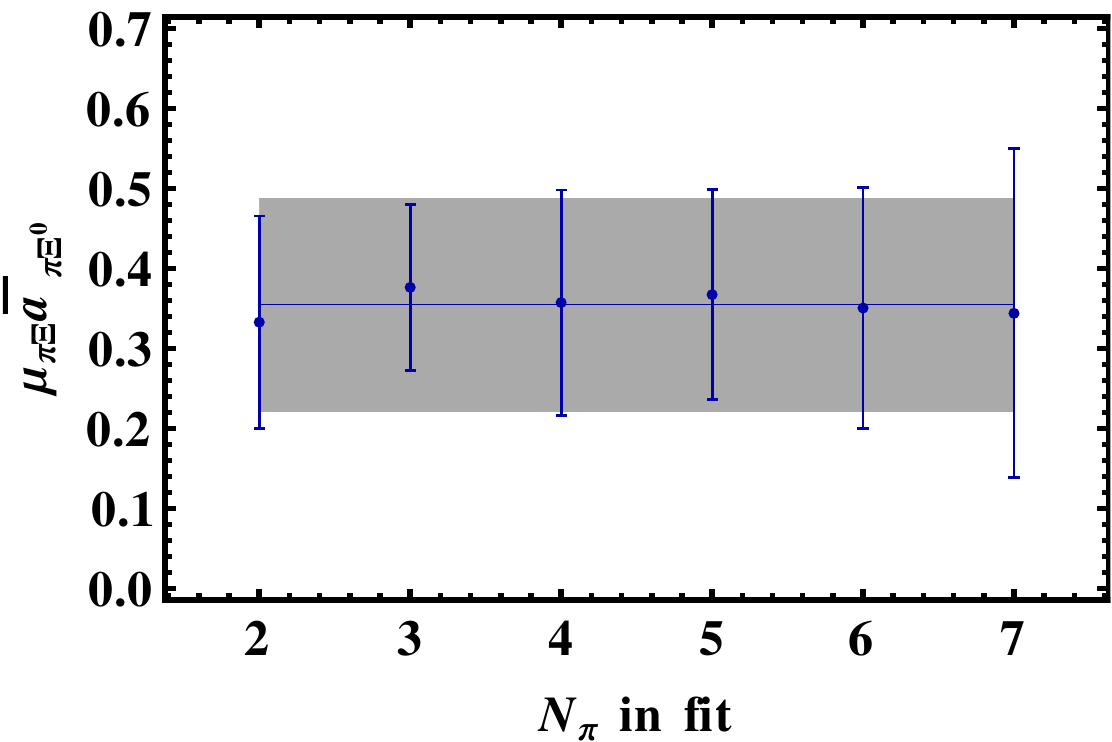}\\
\vspace{1mm}
\includegraphics[width=0.4\linewidth]{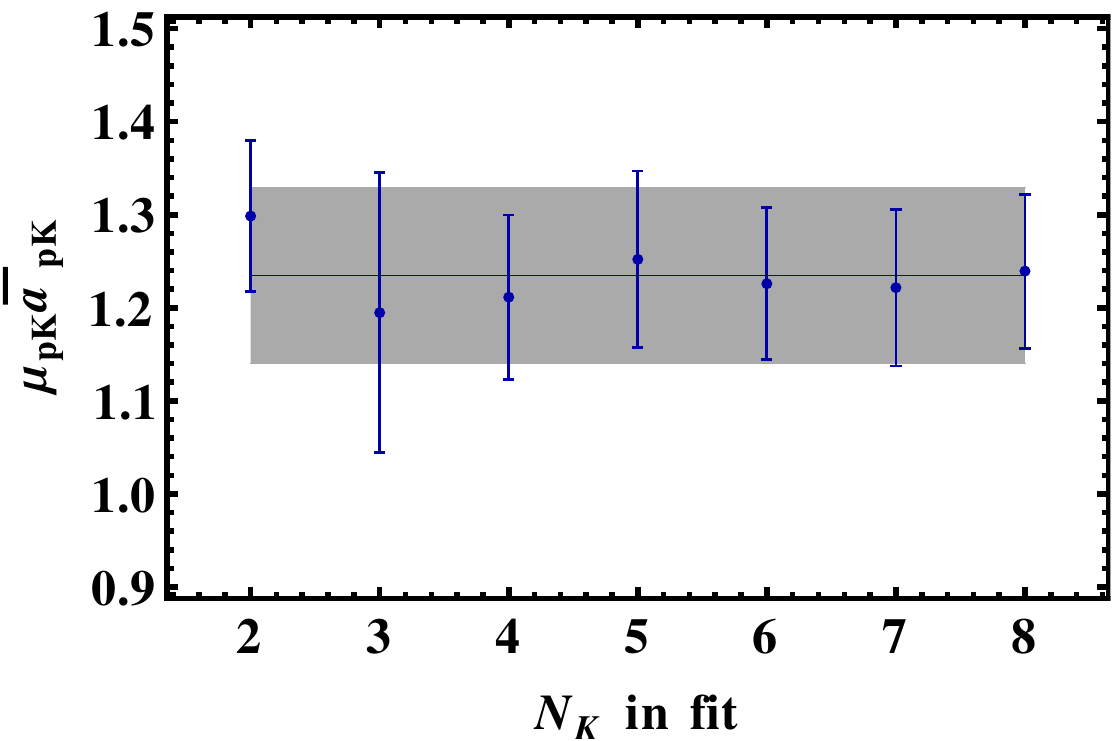}\hspace{2mm}
\includegraphics[width=0.4\linewidth]{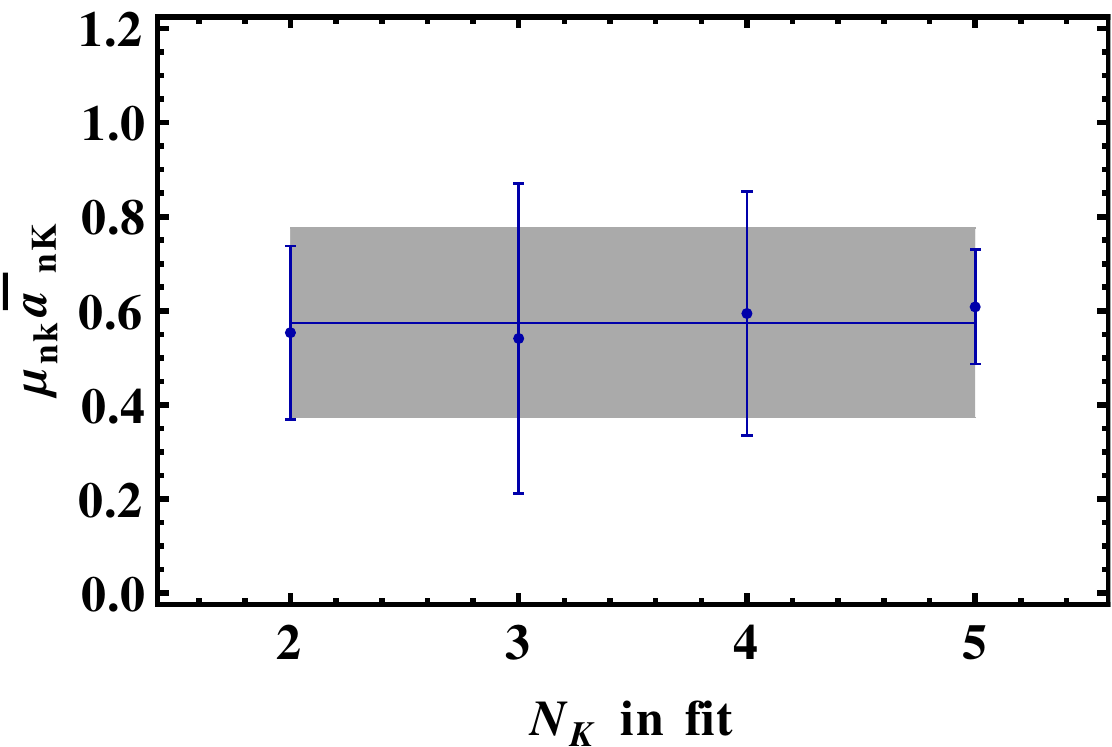}
\caption{\label{fig:aMB}Effective scattering lengths, $\bar{a} = \left(p\cot\delta\right)^{-1}$, as a function of the maximum system size included in the fit, with error bars representing combined statistical and systematic uncertainties. Clockwise from upper left: $\pi^{+}\Sigma^{+}$, $\pi^{+}\Xi^0$, $K^{+}$n, and $K^{+}$p. The gray band shows the mean of all fits and their uncertainties, plus an additional uncertainty given by the standard deviation of all fits, added in quadrature.}
\end{figure}

\begin{figure}
\centering
\includegraphics[width=0.4\linewidth]{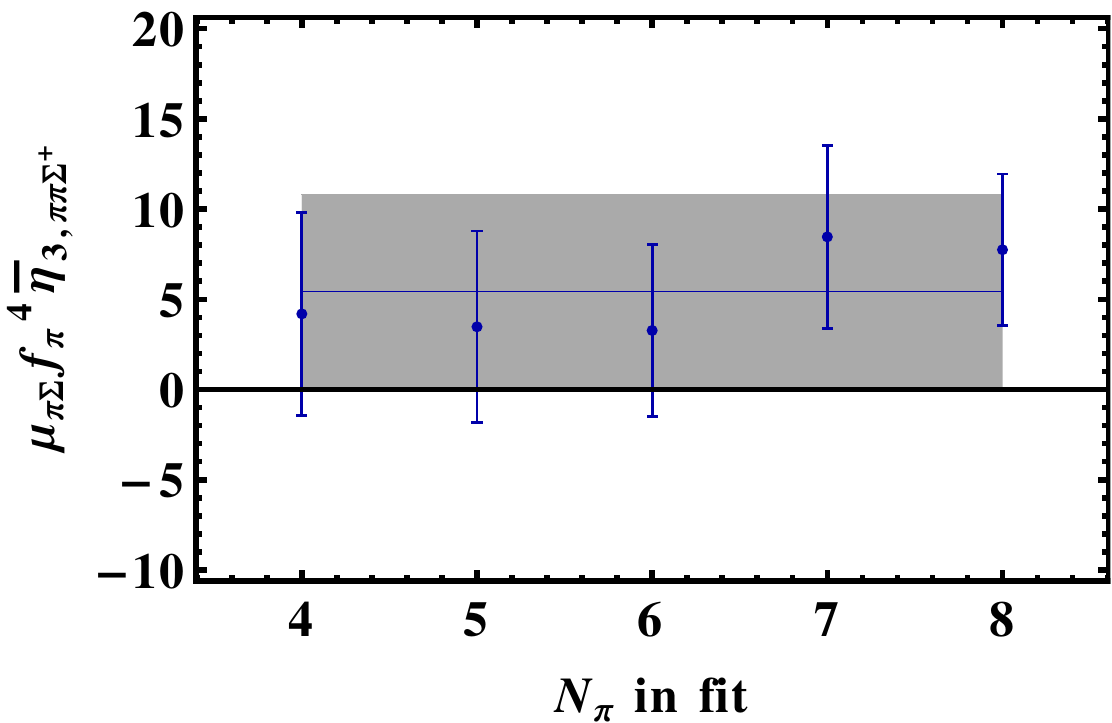}\hspace{2mm}
\includegraphics[width=0.4\linewidth]{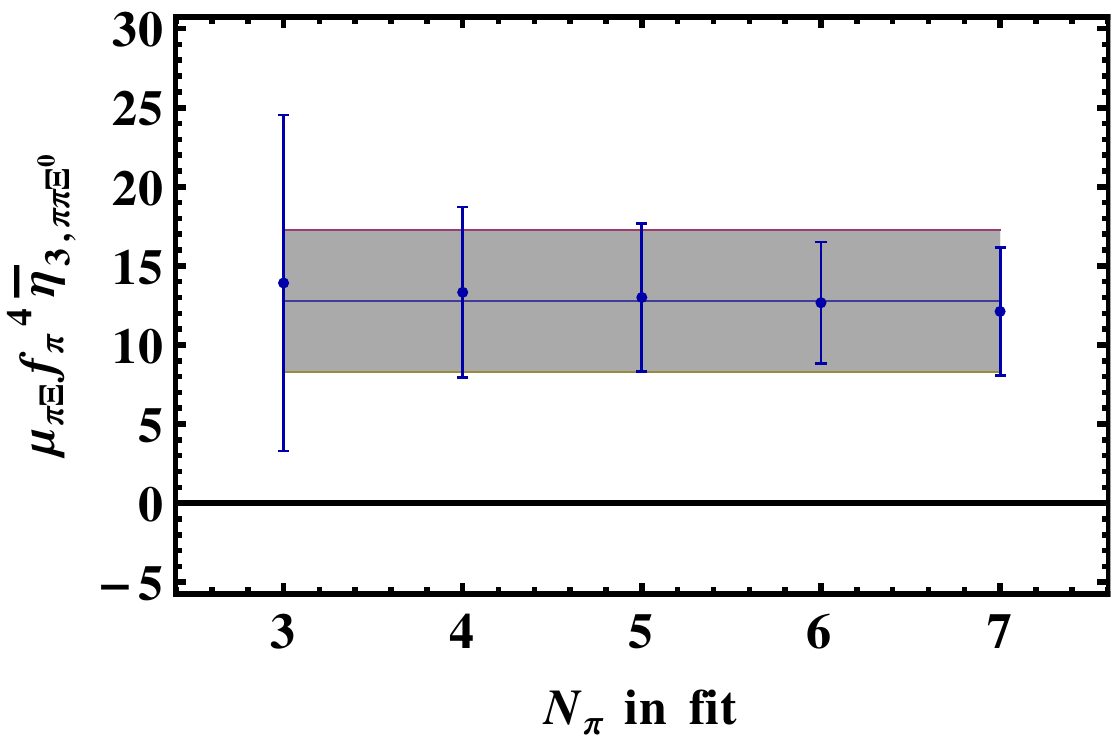} \\
\vspace{1mm}
\includegraphics[width=0.4\linewidth]{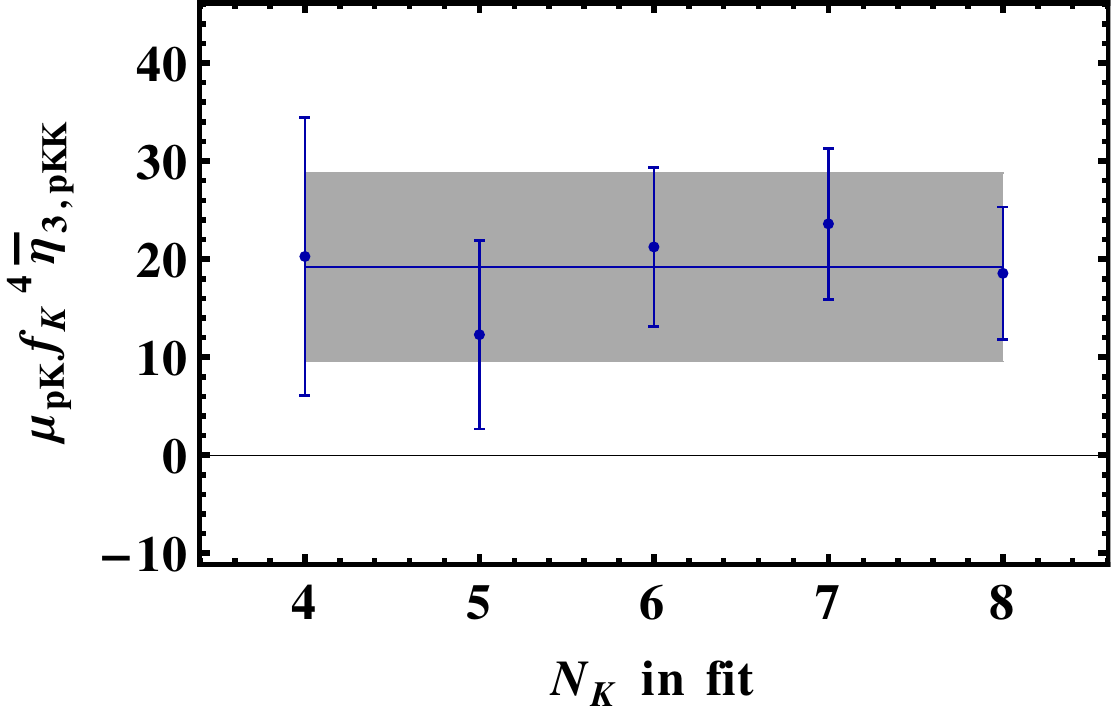}\hspace{2mm}
\includegraphics[width=0.4\linewidth]{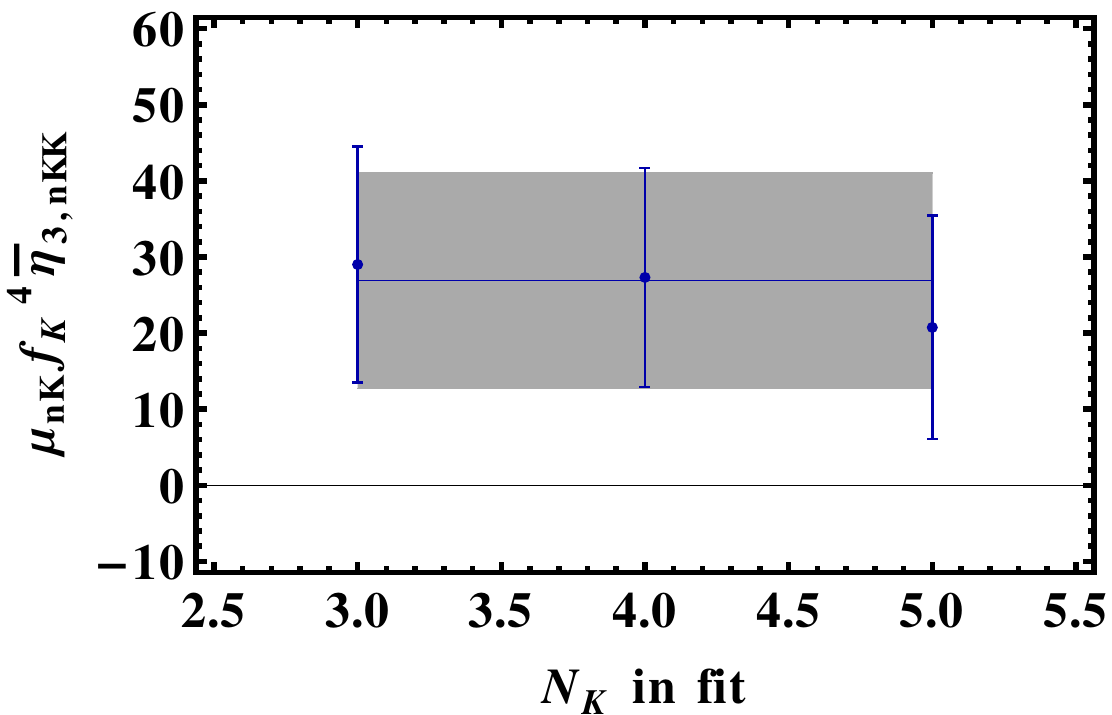}
\caption{\label{fig:etaMMB}Three-body interaction parameters, $\bar{\eta}_{BMM}$, as a function of the maximum system size included in the fit, with error bars representing combined statistical and systematic uncertainties. Clockwise from upper left: $\pi^{+}\pi^{+}\Sigma^{+}$, $\pi^{+}\pi^{+}\Xi^0$, $K^{+}K^{+}$n, and $K^{+}K^{+}$p. The gray band shows the mean of all fits and their uncertainties, plus an additional uncertainty given by the standard deviation of all fits, added in quadrature.}
\end{figure}

Results for the two-body meson-baryon interactions are shown in \Fig{aMB} as a function of the number of mesons included in the fit. We find no significant variation with the system size. However, we do find significant scattering momentum dependence when compared with previous results at a smaller volume from the NPLQCD collaboration \cite{Torok:2009dg}. This likely indicates that contributions from short-distance physics, such as the effective range, $t$-channel cuts, or inelasticities, are relevant at these momenta; understanding these contributions will be the subject of future work. We have also extracted the meson-meson two-body effective scattering lengths and three-meson interactions using the single-species relation (Ref.~\cite{Beane:2007qr,Detmold:2008gh}). We find results consistent with those from several other groups (see Ref.~\cite{Detmold:2013gua}). Finally, in \Fig{etaMMB}, we plot the meson-meson-baryon three-body interactions extracted from our ground state energies using \Eq{LuscherExp}. These are novel results; we find nonzero contributions for most systems within our uncertainties.

\section{Tree-level chiral perturbation theory}

Masses of several low-lying baryons in the presence of a pion condensate have been computed using tree-level HB$\chi$PT \cite{Bedaque:2009yh}. The condensate mixes the baryons, and the masses of the ground-state baryons having the quantum numbers of the vacuum $\Sigma^{+}$ and $\Xi^0$ were found to be,
\beq
M_{\Sigma^{+}}(\mu_I) &=& M_{\Sigma}^{(0)}+4c_1^{\Sigma} m_{\pi}^2\cos \alpha +(c_2^{\Sigma}+c_3^{\Sigma}+c_6^{\Sigma}+c_7^{\Sigma})\mu_I^2 \sin^2\alpha \cr
&-&\mu_I\sqrt{\cos^2\alpha+(c_6^{\Sigma}+c_7^{\Sigma})^2\mu_I^2 \sin^4\alpha} \ , \cr
M_{\Xi^{0}}(\mu_I) &=&  M_{\Xi}^{(0)}-\frac{\mu_I}{2} \cos \alpha +4c_1^{\Xi}m_{\pi}^2 \cos \alpha + \left(c_2^{\Xi}-\frac{g_{\Xi\Xi}^2}{8M_{\Xi}^{(0)}}+c_3^{\Xi}\right) \mu_I^2 \sin^2\alpha \,
\eeq  
where $M_X^{(0)}$ is the mass of the baryon in the chiral and zero chemical potential limits, all LECs, $c_{i}^{X}$, are as defined in \cite{Bedaque:2009yh}, and $\cos \alpha =\{ 1$, $\mpi^2/\mu_I^2\}$, where $\mu_I = \frac{1}{2} (\mu_u-\mu_d)$, in the vacuum and pion condensed phases, respectively. 

We can derive similar relations for the ground-state baryons with the quantum numbers of the nucleons in a kaon condensate using $SU(3)$ HB$\chi$PT. We find,
\beq
M_{n}(\mu_K) &=&M_{n}^{(0)} -\frac{\mu_K}{2} \cos \alpha + \left(2b_0+b_D-b_F\right)m_{K}^2 \cos \alpha \cr
&+&\frac{1}{4}\left(b_1-b_2+b_3+b_4-b_5+b_6+2b_7+2b_8\right)\mu_K^2 \sin^2\alpha\ , \cr
M_{p}(\mu_K) &=& M_{p}^{(0)} +2 \left(b_0+b_D\right)m_{K}^2 \cos \alpha +\frac{1}{2}\left(b_1+b_3+b_4+b_6+b_7+b_8\right)\mu_K^2 \sin^2\alpha\cr
&-&\sqrt{\left(2b_F(m_K^2-m_{\pi}^2)+\mu_K\cos\alpha\right)^2+\frac{1}{4}(b_1-b_3+b_4-b_6)\mu_K^4 \sin^4\alpha} \ ,
\eeq
where the $b_i$ are LECs of the $SU(3)$ HB$\chi$PT Lagrangian, and $\mu_K = \frac{1}{2}(\mu_u-\mu_s)$.

\begin{figure}
\centering
\includegraphics[width=0.4\linewidth]{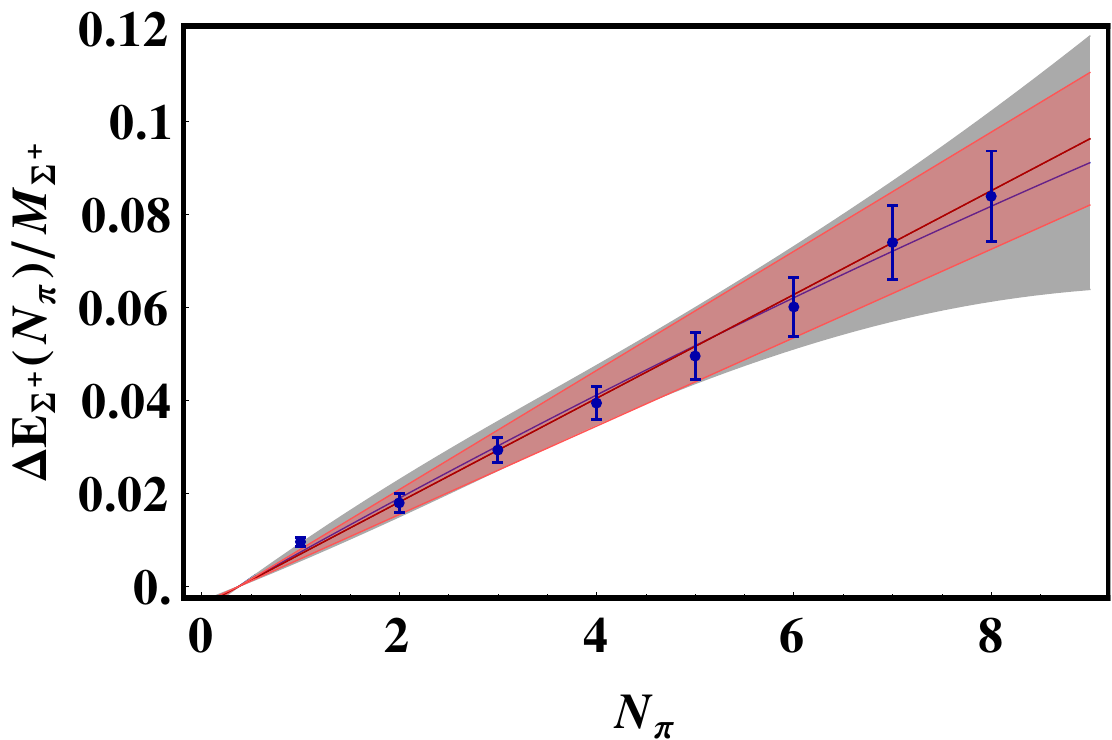}\hspace{2mm}
\includegraphics[width=0.4\linewidth]{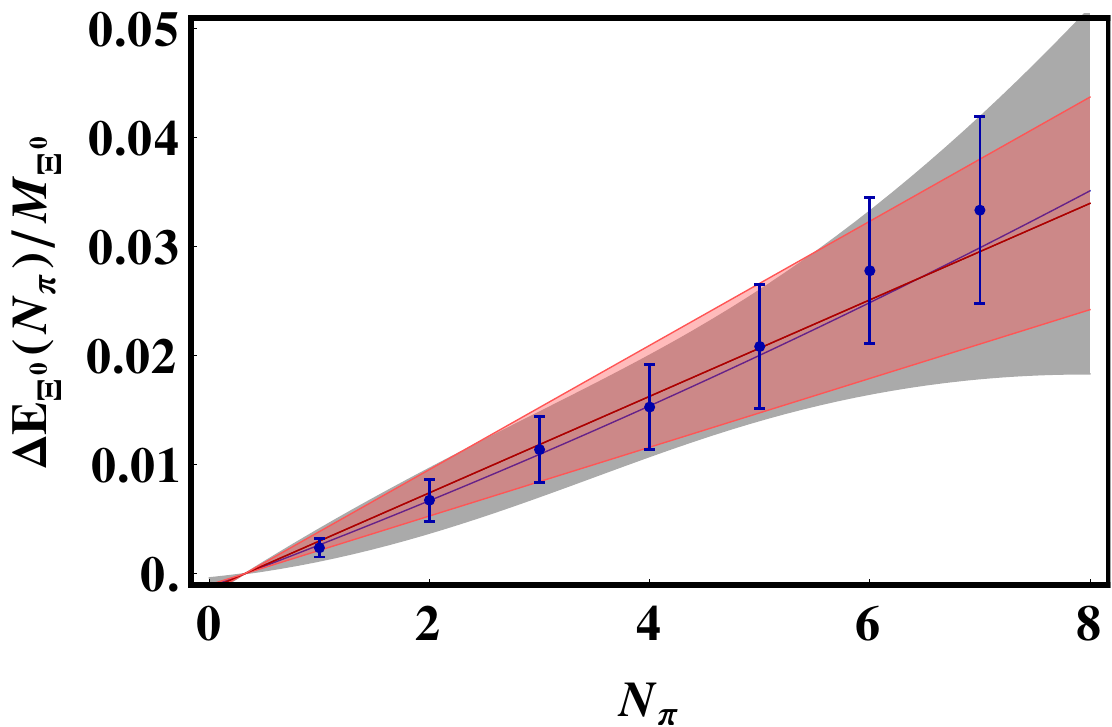} \\
\vspace{1mm}
\includegraphics[width=0.4\linewidth]{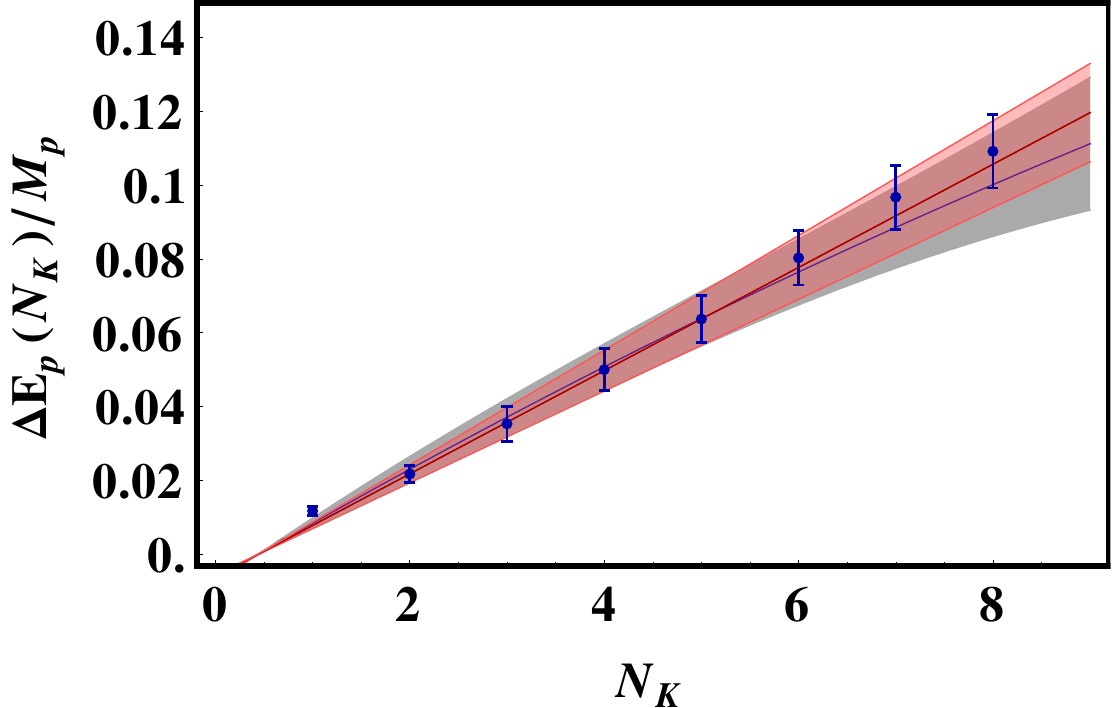} \hspace{2mm}
\includegraphics[width=0.4\linewidth]{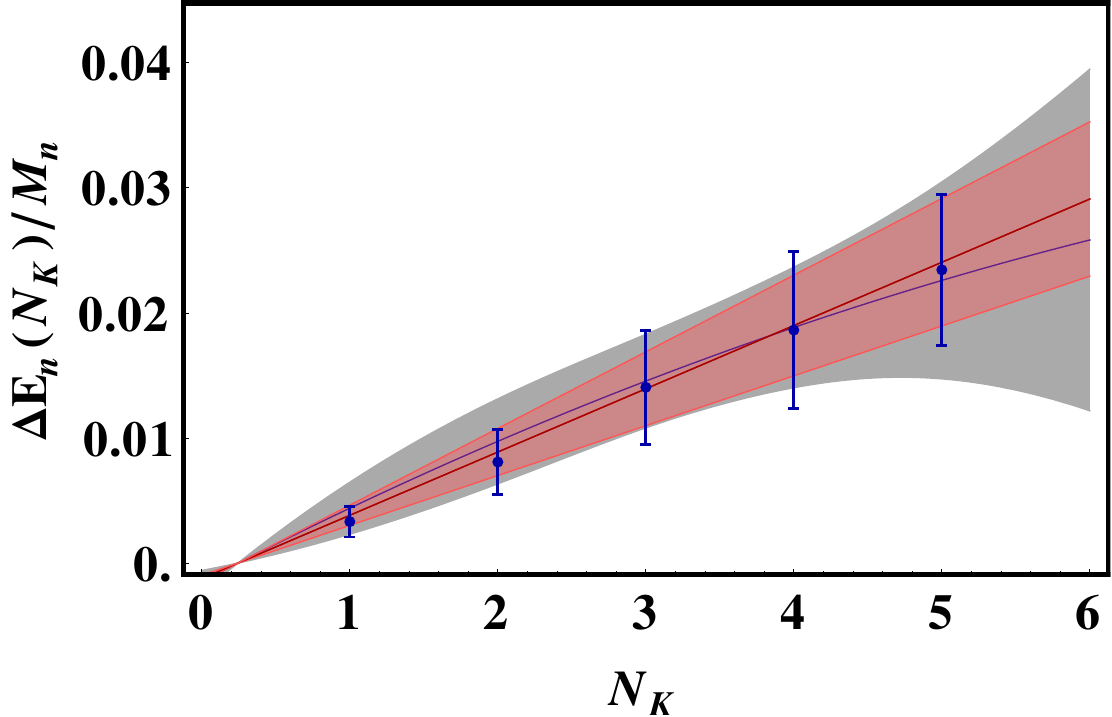}
\caption[]{\label{fig:ChiPT}Energy splittings in units of the baryon mass as a function of the number of mesons, with error bars representing combined statistical and fitting systematic uncertainties. Clockwise from upper left:  $(\pi^{+})^{N_{\pi}}\Sigma^{+}$, $(\pi^{+})^{N_{\pi}}\Xi^0$, $(K^{+})^{N_K}$n, and $(K^{+})^{N_K}$p.  One- and two-parameter fits to \Eq{massExp}, are shown as red and gray shaded bands, respectively.}
\end{figure}

To connect our results using a canonical formulation to the grand canonical relations with an external source for the baryon, we use a finite energy difference to determine the chemical potential, $\mu_{I,K}(n) = E_{\pi,K}(n+1)-E_{\pi,K}(n)$, where $E_{\pi,K}(n)$ is the energy of the system of $n$ pions or kaons, respectively. We must also subtract off the direct coupling of the baryon to the chemical potential in vacuum, $M_{B}(\mu_{I,K},\cos \alpha = 1)$. Finally, we find that the chemical potentials of our systems are very near the condensation point, $\mu_{I,K} \approx m_{\pi,K}$, and thus we use an expansion of the mass splittings about the condensation point to improve our fits,
\beq
\label{eq:massExp}
\Delta M_B/m_{\pi,K} = \left( 2Q_{I,K}+4a^{(B)}\right) \left( \frac{m_{\pi,K}}{\mu_{I,K}}-1\right) + \left(-Q_{I,K}+2 b^{(B)}\right) \left( \frac{m_{\pi,K}}{\mu_{I,K}}-1\right)^2 + \cdots \ ,
\eeq
where $Q_{I,K}$ is the isospin/kaon charge of the baryon, and the combinations of LECs corresponding to the parameters $a^{(B)}$ and $b^{(B)}$ are given in Ref.~\cite{Detmold:2013gua}. Plots of our results for the mass splittings versus the system size with fits to \Eq{massExp} are shown in \Fig{ChiPT}. Fit results for the parameters $a^{(B)}$ and $b^{(B)}$ are given in \cite{Detmold:2013gua}. We are unable to extract significant values for the parameters $b^{(B)}$ given our statistics and chemical potentials. 

\begin{comment}
\begin{figure}
\centering
\includegraphics[width=0.4\linewidth]{Sigma_a.pdf}\hspace{2mm}
\includegraphics[width=0.4\linewidth]{Xi_a.pdf}\\
\vspace{1mm}
\includegraphics[width=0.4\linewidth]{Proton_a.pdf}\hspace{2mm}
\includegraphics[width=0.4\linewidth]{Neutron_a.pdf}
\caption{\label{fig:abSig}One- and two-parameter fit results (red and blue data points, respectively) for $a^{\Sigma}$ as a function of the minimum system size included in the fit, with clusters of points having higher maximum system sizes from left to right, and error bars representing combined statistical and systematic uncertainties. Clockwise from upper left:  $(\pi^{+})^{N_{\pi}}\Sigma^{+}$, $(\pi^{+})^{N_{\pi}}\Xi^0$, $(K^{+})^{N_K}$n, and $(K^{+})^{N_K}$p. The gray band shows the mean of all 1-parameter fits and their uncertainties, plus an additional uncertainty given by the standard deviation of all fits, added in quadrature.}
\end{figure}
\end{comment}

\section{Summary}
We have presented results of a lattice calculation of the ground-state energies of several baryons in the presence of a medium of pions or kaons. From the ground-state energies we have extracted two-and three-body interaction parameters, as well as certain combinations of LECs. We find significant momentum dependence for the meson-baryon scattering phase shifts at momenta much smaller than the pion mass when comparing to previous results from NPLQCD \cite{Torok:2009dg}. We also find novel non-zero meson-meson-baryon three-body interactions. Future work will explore the momentum dependence of the meson-baryon phase shifts, as well as possible modifications of the dispersion relations of the baryons in medium.

\begin{acknowledgments}
The authors would like to thank A. Walker-Loud, R. Brice\~{n}o, B. Smigielski, J. Wasem, P. Bedaque, S. Wallace, M. Savage, and D. Kaplan for helpful discussions. We thank the NPLQCD collaboration for allowing us to use previously generated meson and baryon blocks. WD was supported in part by the U.S. Department of Energy through Outstanding Junior Investigator Award DE-SC000-1784 and Early Career Award DE-SC0010495. AN was supported in part by U.S. DOE grant No. DE-FG02-93ER-40762. 

\end{acknowledgments}

\bibliographystyle{physrev.bst}
\bibliography{Mesonsref}

\end{document}